\documentclass{IEEEcsmag}

\usepackage[colorlinks,urlcolor=blue,linkcolor=blue,citecolor=blue]{hyperref}
\usepackage[table,xcdraw]{xcolor}
\usepackage{amsmath}
\usepackage{amssymb}
\usepackage{upmath}
\usepackage{todonotes}
\usepackage{comment}
\usepackage{subcaption}
\usepackage{caption} 
\usepackage{multirow}
\usepackage{array}
\jvol{XX}
\jnum{XX}
\paper{8}
\jmonth{May/June}
\jname{IEEE Security \& Privacy}
\pubyear{2022}

\newcommand{\loss}{L}

\newcommand{\vct}[1]{\ensuremath{\boldsymbol{#1}}}

\newcommand{\myparagraph}[1]{\vspace*{0.05cm} \noindent \textbf{#1}}

\begin{document}


\title{Practical Attacks on Machine Learning: A Case Study on Adversarial Windows Malware}

\author{~Luca Demetrio}
\affil{~University of Cagliari, Italy, and Pluribus One}

\author{~Battista Biggio}
\affil{~University of Cagliari, Italy, and Pluribus One}

\author{~Fabio Roli}
\affil{~University of Genova, Italy, and Pluribus One}

\markboth{Department Head}{Paper title}

\begin{abstract}
While machine learning is vulnerable to adversarial examples, it still lacks systematic procedures and tools for evaluating its security in different application contexts.
In this article, we discuss how to develop automated and scalable security evaluations of machine learning using practical attacks, reporting a use case on Windows malware detection.
%
\end{abstract}

\maketitle

\chapterinitial{Introduction}

Machine learning has recorded unprecedented success in many applications, including computer vision and speech recognition.
Even in the cybersecurity domain, many companies have recently built machine learning models within their detection pipelines to improve their anti-malware solutions~\cite{demetrio21-tops}.
However, it is now widely known that machine learning models can be easily misled by carefully-crafted attacks, such as training data poisoning, backdooring, evasion, model stealing, and other privacy-related threats~\cite{biggio18}. 
While many of these attacks can be successfully prevented, machine learning models remain extremely vulnerable to \emph{adversarial examples}~\cite{biggio13-ecml,szegedy14-iclr}, that are inputs presented at test time specifically designed to cause the model to make a mistake.
Adversarial examples are normally found by optimizing a perturbation against the target model either via gradient-based optimization, when white-box access to the model is given (the kind of model and its trained parameters are accessible), or via gradient-free optimizers, when only black-box access to the model is provided (for instance, the model can be queried using different inputs, and feedback on the corresponding predictions is observable). 
In the black-box setting, it is also possible to stage \textit{transfer attacks}, which are gradient-based attacks optimized against a surrogate model which also succeed against the target model.
Such attacks are feasible only when the surrogate model provides a differentiable and sufficiently-smooth approximation of the target model, which is clearly neither always available to the attacker nor easy to build~\cite{biggio18}.
\begin{figure}
    \centering
    \includegraphics[width=\linewidth]{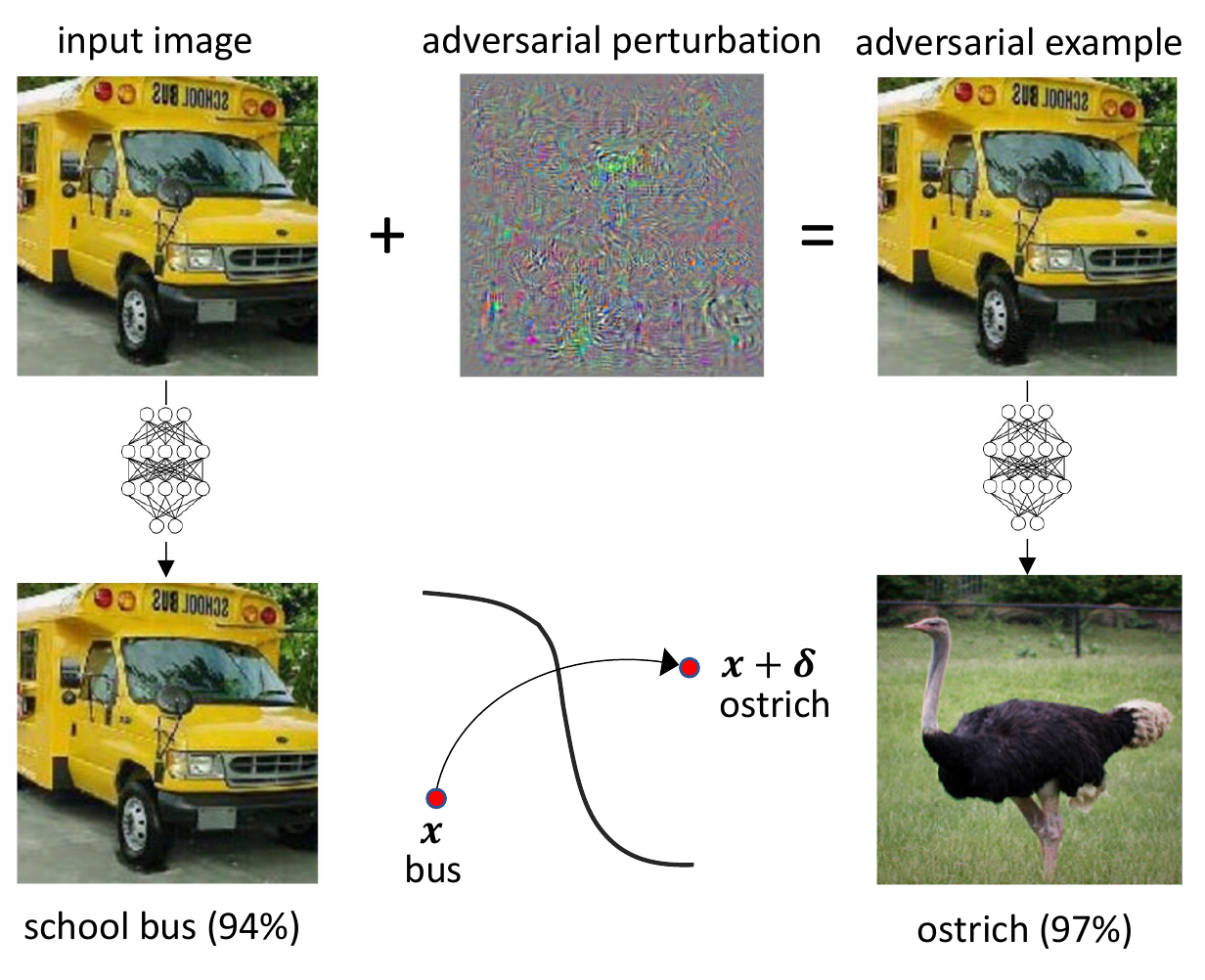}
    \caption{Adversarial examples are crafted by optimizing an input perturbation $\vct \delta$ to fool the target model. In this example, a slightly-perturbed image of a school bus is misclassified as an ostrich.}
    \label{fig:adv_examples}
\end{figure}
In Figure~\ref{fig:adv_examples}, we exemplify the process used to craft adversarial examples, starting from the image of a school bus (classified correctly with 94\% confidence by a state-of-the-art model trained on ImageNet), and showing how it can be perturbed to generate an adversarial example misclassified as an ostrich (with 97\% confidence).
The latter is computed by solving the following optimization problem:
\begin{equation}
    \begin{aligned}
        \max_{\vct \delta} & \quad \loss(\vct x + \vct \delta, y; \vct \theta),\\
        \rm{s.t.} & \quad  \| \vct \delta \|_p \leq \epsilon, \quad \vct x + \vct \delta \in [0,1]^d,
    \end{aligned}
    \label{eq:img_adv}
\end{equation}
where $L(\vct x, y, \theta)$ is a loss function that exhibits lower values when the input sample $\vct x$ (for instance, the input image consisting of $d$ pixels) is correctly assigned to class $y$ by the model (parameterized via $\vct \theta$). 
The goal is to optimize the applied perturbation $\vct \delta$, to maximize the loss on the perturbed sample $\vct x + \vct \delta$, to produce a misclassification with high confidence by the target model.
However, this should be achieved while preserving some constraints on the applied perturbation. 
First, the $\ell_p$ norm of $\vct \delta$ is typically upper bounded by a small number $\epsilon$, to keep the perturbation size small. 
Second, the perturbed sample $\vct x + \vct \delta$ is also normally constrained to stay within some bounds, e.g., to ensure that each pixel of the perturbed image lies in the scaled interval $[0,1]$.
While this formulation works well for the image domain, it is not straightforward to extend it to other domains.
First, adding perturbations to the input data is not acceptable in many applications; 
for instance, crafting adversarial malware requires manipulating complex structures in the input program, which can not be formalized as additive perturbations. 
Second, constraining the perturbation size using $\ell_p$ norms may not have any practical meaning for the application at hand.

These issues hinder the applicability and the generality of these attacks beyond the image domain, highlighting the need for more general and \emph{practical} adversarial attacks.
In particular, we believe that, for machine learning attacks to become \textit{practical}, the considered threat models have to satisfy four main properties, that ensure the corresponding input perturbations have to be:
\begin{enumerate}
    \item \emph{application-specific}, as they should enable crafting real-world attacks in different applications (e.g., image manipulations are different from perturbations which can be applied to source code);
    \item \emph{semantics-preserving}, as they must comply with constraints imposed by the given application domain, not to compromise the content or functionality of the source input samples; 
    \item \emph{automatable}, as the process of crafting attacks should be repeatable and scalable, without requiring extensive human intervention, and 
    \item \emph{fine-tunable}, as they should ensure good testing coverage of the input space to also identify adversarial examples lying in hard-to-find blind spots.
\end{enumerate}

In this work, we show that adversarial attacks can be generalized to encompass more complex, practical, and application-specific manipulations.
We develop a unifying framework for computing adversarial attacks that parameterizes these manipulations, enabling the production of minimal, content- and functionality-preserving adversarial examples.
We present a case study on Windows malware detection by systematizing and defining all the known feasible manipulations that abuse the PE file format flexibility to craft evasive malware. 
We show how the corresponding attacks can highly degrade the performances of popular machine learning-based malware detectors under both white-box and black-box attack scenarios, and how these attacks also surprisingly transfer to some well-known commercial products.

We conclude this article by envisioning the creation of more security tools for the various domains where machine learning is applied, followed by integrated development environments for creating, maintaining, versioning, debugging, and testing machine learning models before their deployment. 
We firmly believe that this will be a remarkable step towards bringing the current machine learning development practices much closer to the best practices which are normally followed in modern software engineering, easing deployment, maintainability, testing, and security of machine learning models in real-world applications.

\section{Practical Attacks against Machine Learning}

\begin{figure}
    \centering
    \includegraphics[width=\linewidth]{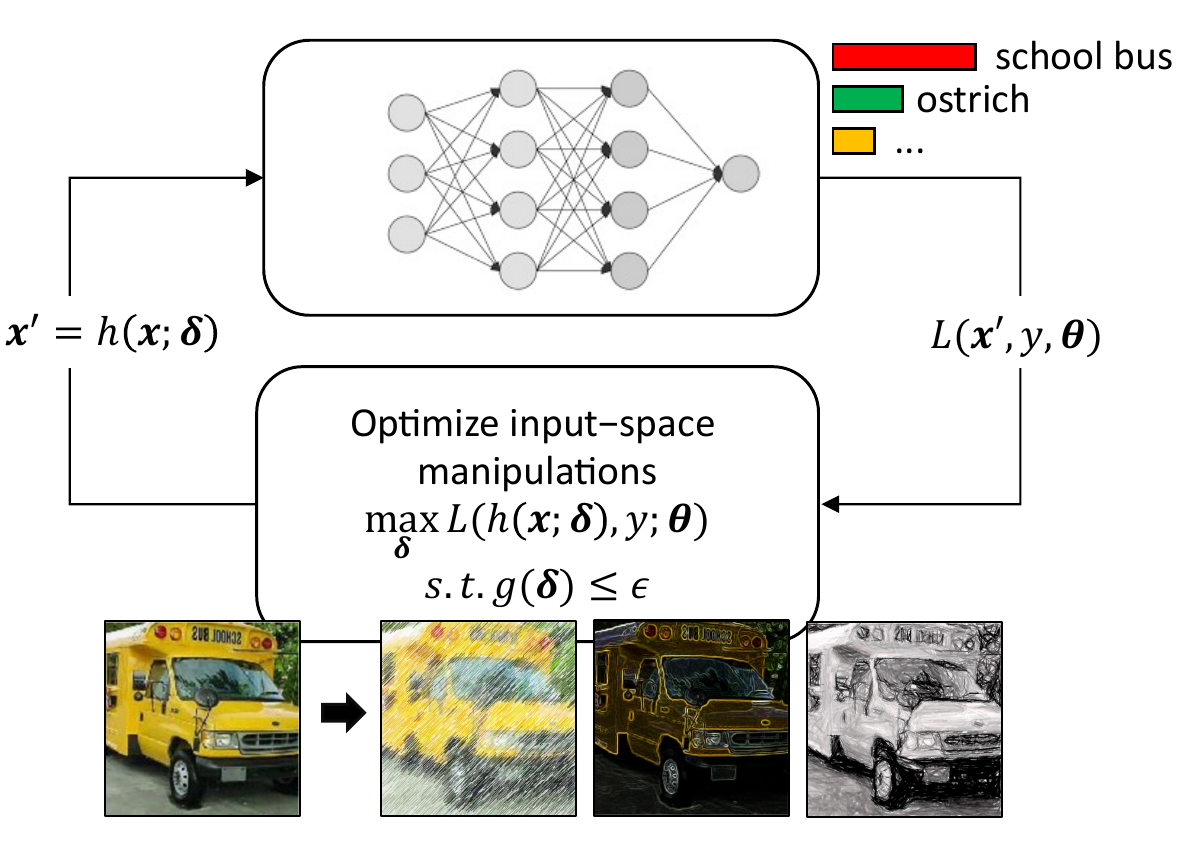}
    \caption{Our framework builds upon two core components: parametric manipulations customized for the application domain and an optimization algorithm. Hence, each attack applies manipulations, and the optimizer adjusts the size of the perturbation that is applied to an input sample at each step of the strategy, ensuring that the given constraints are satisfied.}
    \label{fig:framework}
\end{figure}

We discuss here how to overcome the four factors that are hindering the development of large-scale, practical security attacks of machine learning in different application-specific contexts. 
To this end, we envision a practical framework consisting of two main building blocks: (i) a set of practical, application-specific manipulations that can be applied to craft perturbed input samples; and (ii) an optimization algorithm that identifies the best combination of such manipulations to find the corresponding adversarial examples, by also considering an application-specific function to bound the perturbation size.
We conceptually represent this two-step procedure in Figure~\ref{fig:framework}, and provide below a more detailed description of each step.

\subsubsection{Practical Manipulations.} 
As anticipated, it is important to define a set of feasible manipulations based on the properties of the input data which have to be perturbed.
Such manipulations have thus to be \emph{application-specific}, meaning that each domain should be investigated in detail to understand how to implement perturbation models that are well suited to the given input data.
In the proposed framework, we model the set of feasible manipulations via a manipulation function $h$, parameterized by a vector $\vct \delta$, hence $h(\vct x; \vct \delta)$ creates a perturbed version of the input sample $\vct x$.
The underlying idea is to use the parameter vector $\vct \delta$ to control and optimize the type and intensity of the applied perturbation; for example, if we assume that $h$ corresponds to manipulating images by rotating them, then $\delta$ may simply be a scalar value corresponding to the degrees of rotation.
This also ensures that the manipulations are \emph{fine-tunable}, implying that they can be optimized against a given target model.
Finally, the manipulation function $h$ must also be \emph{semantics-preserving}, as it must preserve the semantics of the perturbed object by design to ensure that the content or functionality of the input data remains intact; for instance, adversarial malware must preserve its malicious functionality while being undetected, and modified spam emails must still convey the intended message to the targeted users while evading anti-spam filters.
We will provide more details and examples of practical manipulations on Windows malware in our case study.

\subsubsection{Attack Optimization.} 
We now describe the second component of the proposed framework by detailing the optimization step.
We hence write a similar optimization problem to Equation~\eqref{eq:img_adv}, including the manipulation function $h$:
\begin{equation}
    \begin{aligned}
       \max_{\vct \delta} & \quad \loss(h(\vct x; \vct \delta), y; \vct \theta),\\
        \rm{s.t.} & \quad g(\vct \delta) \leq \epsilon
    \end{aligned}
    \label{eq:adv}
\end{equation}
where $g$ is an abstraction of the constraint we described in Equation~\eqref{eq:img_adv}, and it can be customized for the target domain.
This optimization problem can be solved using two different families of algorithms, depending on whether white-box or black-box access to the target model is provided. These techniques are normally referred to respectively as (i) gradient-based and (ii) gradient-free optimizers~\cite{biggio18,demetrio21-tops}.

\emph{Gradient-based optimizers} are most suitable when perfect knowledge of the target model is available (i.e., white-box access is provided), and the model is differentiable. 
Thus, perturbations can be iteratively optimized using the information retrieved from its \emph{gradients}.
For instance, this is the case of end-to-end attacks against image classifiers, where gradients are used to drive the optimization of the pixel values towards the desired class. 
Even if the model is trained on handcrafted (non-differentiable) features extracted from the input sample, gradient-based attacks can be still used, provided that the perturbations applied to the input features can be then implemented in practice, by ensuring that one can build the adversarial example corresponding to the perturbed feature representation.

\emph{Gradient-free optimizers} are most suitable when either the model under attack is not differentiable, or only responses to input queries can be retrieved from it, while no knowledge of its internal parameters is available (i.e., only black-box access to the model is given).
Since the target can be queried, these strategies maximize the loss by fine-tuning the manipulations based on the returned predictions.
Alternatively, manipulations can also be optimized via \emph{transfer attacks}, if a surrogate, differentiable model which well approximates the target is available. In this case, gradient-based attacks can be optimized against the surrogate model and then transferred to the target.

All of these families of optimizers allow attacks to be applied automatically, hence adversarial examples can be computed on a large scale, satisfying the \emph{automatable} desired property.

\begin{table*}[ht]
\centering
\begin{tabular}{ccccccc}
\textbf{} &
  \cellcolor[HTML]{C0C0C0}\textbf{Image classification} &
  \cellcolor[HTML]{C0C0C0}\textbf{Speech-to-text} &
  \cellcolor[HTML]{C0C0C0}\textbf{Spam detection} \\
\cellcolor[HTML]{C0C0C0}\textbf{Proposed by} &
  \begin{tabular}[c]{@{}c@{}}Biggio et al.~\cite{biggio13-ecml}\\ Szegedy et al.~\cite{szegedy14-iclr}\end{tabular} &
  Carlini et al.~\cite{carlini2018audio} &
  \begin{tabular}[c]{@{}c@{}}Nelson et al.~\cite{nelson08}\\ Dalvi et al.~\cite{dalvi04}\end{tabular} \\
\cellcolor[HTML]{C0C0C0}\textbf{Manipulations} &
  Additive noise &
  Additive noise &
  \begin{tabular}[c]{@{}c@{}} GWI/BWO \end{tabular} \\
\cellcolor[HTML]{C0C0C0}\textbf{Optimizer} &
  Gradient-based &
  Gradient-based &
  Gradient-based \\
\cellcolor[HTML]{C0C0C0}\textbf{Constraint} &
  $\ell_2, \ell_\infty$ &
  $\ell_2$ &
  $\ell_0$ \\
\textbf{} &
  \cellcolor[HTML]{C0C0C0}\textbf{Windows malware detection} &
  \cellcolor[HTML]{C0C0C0}\textbf{PDF malware detection} &
  \cellcolor[HTML]{C0C0C0}\textbf{Android malware detection} \\
\cellcolor[HTML]{C0C0C0}\textbf{Proposed by} &
  Demetrio et al.~\cite{demetrio21-tifs, demetrio21-tops} &
   \begin{tabular}[c]{@{}c@{}}
   Biggio et al.~\cite{biggio13-ecml} \\
   Maiorca et al.~\cite{maiorca19-csur}\end{tabular} &
  \begin{tabular}[c]{@{}c@{}}Demontis et al.~\cite{demontis2017yes}\\ Grosse et al.~\cite{grosse2017adversarial}\end{tabular} \\
\cellcolor[HTML]{C0C0C0}\textbf{Manipulations} &
  Format ambiguities &
  Object injection &
  \begin{tabular}[c]{@{}c@{}} Injecting fake APIs, permissions\end{tabular} \\
\cellcolor[HTML]{C0C0C0}\textbf{Optimizer} &
  \begin{tabular}[c]{@{}c@{}}Hybrid optimizer\\ Gradient-free (Genetic)\end{tabular} &
  Gradient-based &
  Gradient-based \\
\cellcolor[HTML]{C0C0C0}\textbf{Constraint} &
  Levenshtein distance &
  Manhattan distance &
  $\ell_0$ \\
\end{tabular}
\caption{Recasting previously-proposed attacks from different application domains (in \textit{columns}) in our framework, detailing the corresponding manipulations and optimizers (in \textit{rows}).
}
\label{table:manipulations_and_optim}
\end{table*}

\subsubsection{Application Examples.}
We discuss here how previously-proposed attacks can be recast into our framework by detailing the considered manipulations, optimizers, and which function they use to constrain the perturbation size.
The analysis for some selected attacks is compactly reported in Table~\ref{table:manipulations_and_optim}.
While images and audio can be manipulated by adding an $\ell_p$-norm bounded perturbation, different kinds of input data require the development of more specific perturbation models. 
For instance, two popular manipulations used to fool anti-spam filters are normally referred to as Good-Word-Injection (GWI) and Bad-Word-Obfuscation (BWO) attacks. 
They respectively consist of modifying a spam email by inserting randomly-chosen words which are likely to appear in legitimate messages but not in spam and by obfuscating (e.g., by misspelling) typical ``spammy'' words.
Similarly, for both Android and Windows malware, the attacker can only inject content by following the conventions imposed by the format used for storing programs as a file.
While GWI and BWO attacks can be constrained using, e.g., the $\ell_0$ norm, to bound the number of modified or injected words, crafting adversarial malware may require defining additional application-specific constraints to bound the perturbation size (e.g., defining distances between sequences of bytes), thus going beyond additive perturbation models.
Recall also that, contrary to the other domains, the misplacement of a single byte in the input program will most likely result in the corruption of the whole executable.

\section{Adversarial Attacks against Windows Malware Detection}
\label{sec:pe}

We discuss here an implementation of our framework, presenting a detailed use case on practical attacks against machine learning Windows malware detectors.
This domain is characterized by several constraints, and it requires extra care when manipulating files, as one single misplaced value can break the entire structure and functionality of a program.
Hence, the manipulations must take into account the rigid structure of the Portable Executable (PE) file format, which dictates how programs are stored as files.

\subsubsection{Programs as Files.} The \emph{PE file format} is made up of several headers, followed by
the program code, the initialized constants, and the program resources.
The headers are three: the \emph{DOS Header}, the \emph{PE Header}, and the \emph{Optional Header}.
The \emph{DOS header} is kept for retro-compatibility with the outdated DOS environment, and it also contains code that will print an error message if a user tries to execute such a program into an older version of Windows.
The few important bytes are the magic number \texttt{MZ} at the beginning of the file and the 4-bytes-long value at offset \texttt{0x3c} that points to the beginning of the real header, the \emph{PE header}.
It starts with the \texttt{PE} signature, and this header specifies the characteristics of the file and the size of the last header of the format, that is the \emph{Optional Header}.
This last header, which is not optional, contains most of the relevant information needed by the operating system to properly load the program in memory and execute it.
These headers are followed by \emph{sections}, constructed by two key components: (i) a \emph{section entry} that specifies where to find the content inside the file through the usage of an offset, and (ii) the \emph{section content} itself.

\begin{figure}
    \centering
    \includegraphics[width=0.8\linewidth]{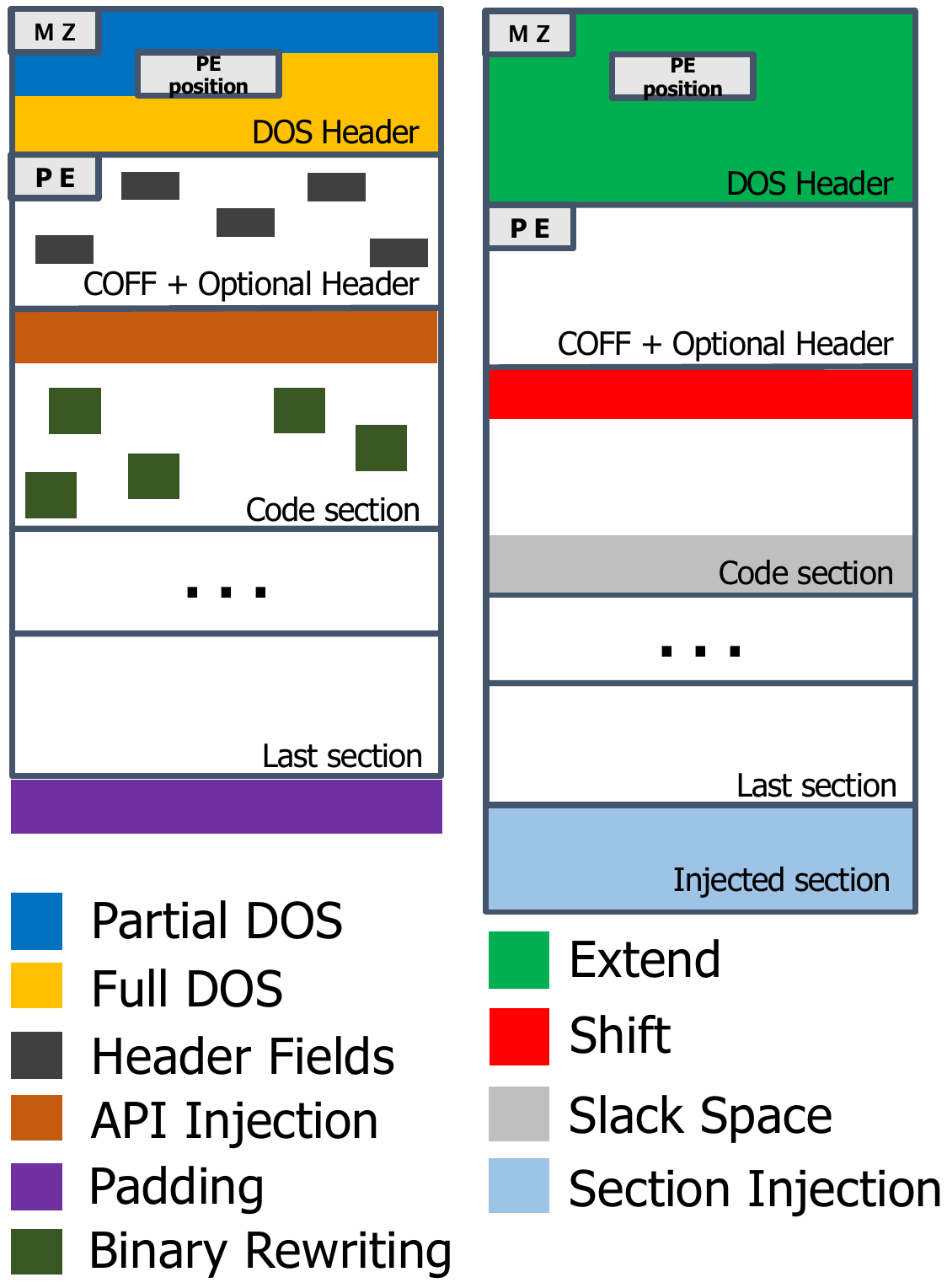}
    \caption{A graphical representation of the PE file format and its manipulations.}
    \label{fig:pe_manipulations}
\end{figure}

\subsubsection{Practical Manipulations of PE Files.} Once the format is known, we dive into the practical manipulations that can be applied safely on a Windows executable without compromising its functionality~\cite{demetrio21-tifs}, as shown in Figure~\ref{fig:pe_manipulations}, where we overlap a graphical representation of the PE file format with its perturbations.
The \emph{Partial DOS} and \emph{Full DOS} manipulations exploit the presence of the useless DOS header in each executable, partially or completely rewriting its unused content.
The \emph{Extend} manipulation leverages the offset that instructs the loader where to find the PE header inside the file by enlarging it and thus reserving space for injecting adversarial content.
The \emph{Header Fields} manipulation perturbs metadata that is not checked by the loader while transferring the content of the program in memory.
The \emph{Shift} manipulation creates space for adversarial content by enlarging the offset of section entries, forcing the loader to look up for each section content further ahead inside the file.
The \emph{Section Injection} manipulation creates and implants a new section entry along with its section content inside the file, providing new space for storing adversarial content.
The \emph{API Injection} manipulation forces the loader to import more functions when uploading the program into memory.
The \emph{Slack Space} manipulation inserts adversarial content inside unused space located between contiguous sections.
The \emph{Padding} manipulation just appends bytes at the end of the sample.
Lastly, we also mention a \emph{Binary Rewriting} techniques~\cite{lucas2021malware} that allows manipulating the program source code by changing instructions or by adding dead code that will never be executed.

\subsubsection{Optimizers for PE Manipulations.}
We now introduce the optimizers that can be used in this domain, depending on the differentiability and the accessibility of the target model to attack.
We summarize these algorithms in Table \ref{tab:optimizers}, along with the attacks built around them.

\begin{table*}[]
\resizebox{0.99\textwidth}{!}{%
\begin{tabular}{cccclc}
 &
  \cellcolor[HTML]{C0C0C0}\textbf{Proposed by} &
  \cellcolor[HTML]{C0C0C0}\textbf{Practical Manipulation} &
  \cellcolor[HTML]{C0C0C0}\textbf{Optimizer} &
  \cellcolor[HTML]{C0C0C0}{\color[HTML]{000000} \textbf{Constraint}} &
  \cellcolor[HTML]{C0C0C0}\textbf{Needs Sandbox} \\
\cellcolor[HTML]{C0C0C0} &
  Demetrio et al. &
  \begin{tabular}[c]{@{}c@{}}Partial DOS, Full DOS, Extend,\\ Shift, Padding, Section Injection\end{tabular} &
  Iterative Discrete Gradient Step &
  Levenshtein distance &
  x \\
\cellcolor[HTML]{C0C0C0}                                          & Kreuk et al.  & Padding, Slack Space & Single Gradient Step & $\ell_2, \ell_\infty$ & x \\
\multirow{-3}{*}{\cellcolor[HTML]{C0C0C0}\textbf{Gradient-based}} & Lucas et al. & Code Rewriting       & Gradient Alignment   & Levenshtein distance                 & x \\ \cline{2-6} 
\cellcolor[HTML]{C0C0C0} &
  Demetrio et al. &
  \begin{tabular}[c]{@{}c@{}}Partial DOS, Full DOS, Extend,\\ Shift, Padding, Section Injection\end{tabular} &
  \begin{tabular}[c]{@{}c@{}}Genetic Optimizer with benign content\\ Transfer\end{tabular} &
  Levenshtein distance &
  x \\
\multirow{-2}{*}{\cellcolor[HTML]{C0C0C0}\textbf{Gradient-free}} &
  Anderson et al. &
  \begin{tabular}[c]{@{}c@{}}Header Fields, API Injection\\ Section Injection, Padding\end{tabular} &
  Reinforcement Learning Agent &
  None &
  \checkmark
\end{tabular}}
\caption{List of algorithms used to attack Windows malware detectors, divided into gradient-based and gradient-free techniques. We also report the manipulations used by each attack, and if they need to validate the created adversarial malware inside a sandbox.}
\label{tab:optimizers}
\end{table*}

When the model under attack is fully accessible and differentiable, \emph{gradient-based} techniques can be used to compute adversarial attacks.
Kreuk et al.~\cite{kreuk2018deceiving} use a single-step approach that uses gradients to manipulate single bytes inside samples.
Since they attack an end-to-end model that abstracts bytes to an embedding layer, once they have computed the perturbation, they need to match it with all the manipulated bytes inside the input space.
However, this method is limited by design, as it only performs one optimization step, thus not exploring the space extensively.
Lucas et al.~\cite{lucas2021malware} apply manipulations that best align with the information provided by the gradient.
They apply random manipulations that alter the code of an executable, and hence they require many iterations to find suitable ones that decrement the malicious score.
Lastly, Demetrio et al.~\cite{demetrio21-tops} apply an iterative optimization algorithm that alters the malware byte-per-byte, thus substituting each selected byte with the closest one that mostly decreases the malicious confidence.
While this technique could in principle require a large number of parameters to tune (one per byte), it is more precise as it iteratively alters all bytes by following the direction pointed by the gradient.

Otherwise, when the model under attack is non-differentiable, or it can only be interacted through queries, \emph{gradient-free} techniques can be used to compute adversarial attacks.
Demetrio et al.~\cite{demetrio21-tifs} use a genetic algorithm to discover the space of manipulations that are applied to malware.
To speed up the process, they inject content extracted from goodware samples, guiding the optimizer towards the benign class.
This formulation avoids the optimization of every single byte of the adversarial noise, hence reducing the number of queries sent to the detector.
Also, they construct transfer attacks against commercial solutions by recycling the adversarial examples computed against the local target.
Anderson et al.~\cite{anderson2017evading} deploy a learning agent that explores the space of manipulations by receiving a reward when it achieves evasion, hence learning which is the best sequence of manipulations to apply.
However, this method not only requires thousands of queries for both training the agent and subsequently evading the target detector, but also the optimizer can break the executable in the process, forcing it to validate the malware inside a sandbox at each iteration of the algorithm.

\myparagraph{Measuring the Perturbation Size.} To bound the manipulation, in this domain, we count the number of modified and injected bytes inside the adversarial malware~\cite{demetrio21-tifs}.
Since programs are represented as strings of bytes, this is equivalent to applying the Levenshtein distance.
Given two strings as inputs, it measures the minimum number of characters that should be inserted, deleted, or substituted in the first string to match the second one.

\section{Experimental Analysis}
\begin{figure*}
\begin{subfigure}{.49\textwidth}
    \centering
    \includegraphics[width=\linewidth]{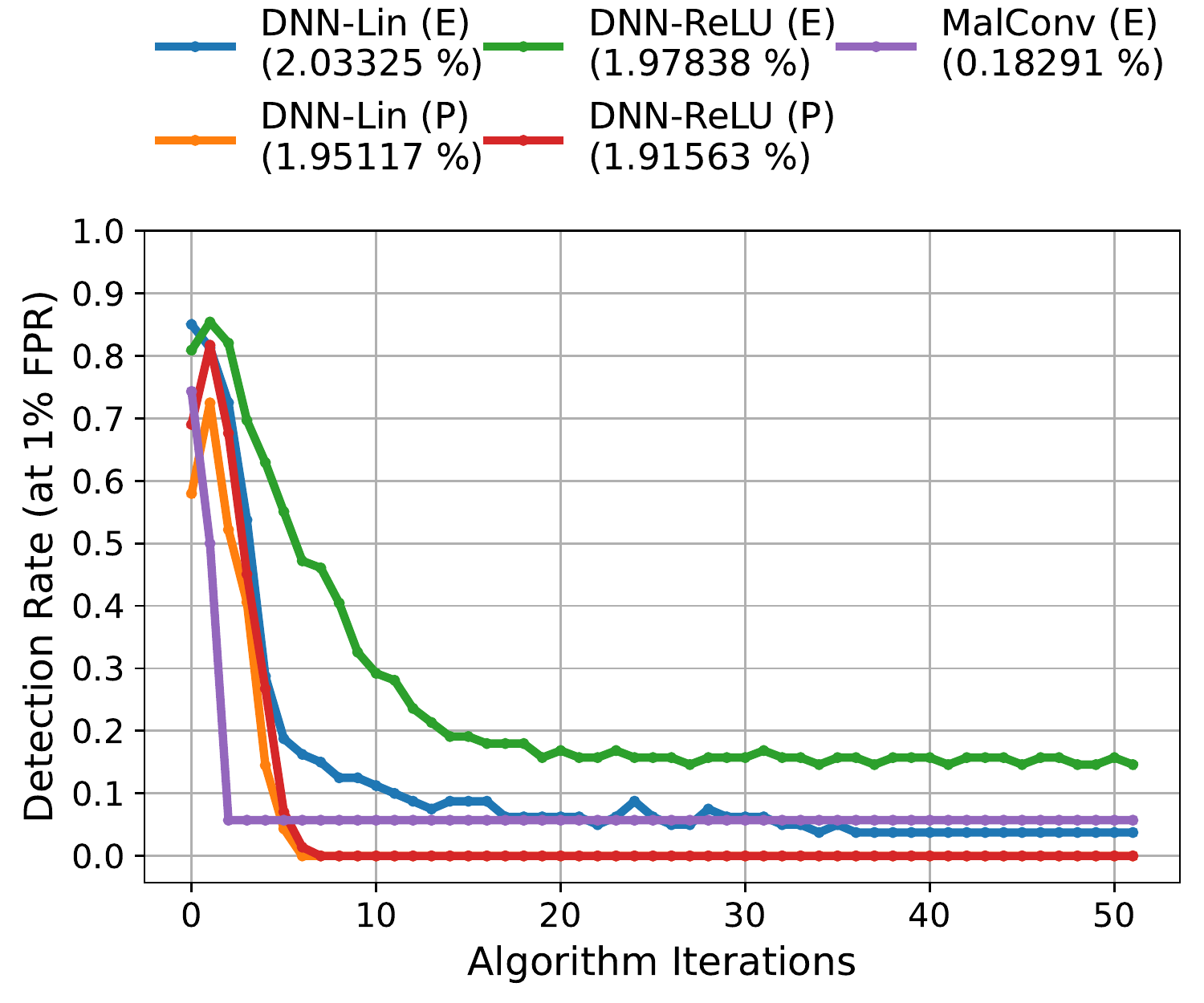}
    \caption{}
    \label{fig:wb_attack}
\end{subfigure}%
\begin{subfigure}{.49\textwidth}
    \begin{subfigure}{\linewidth}
        \centering
        \includegraphics[width=\linewidth]{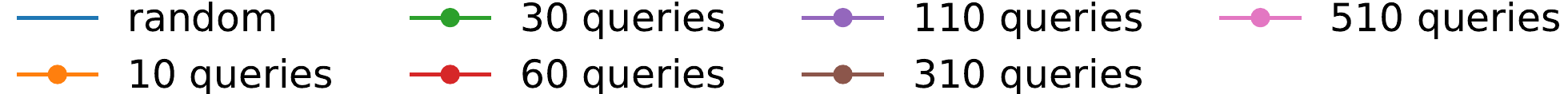}
    \end{subfigure}
    \centering
    \includegraphics[width=\linewidth]{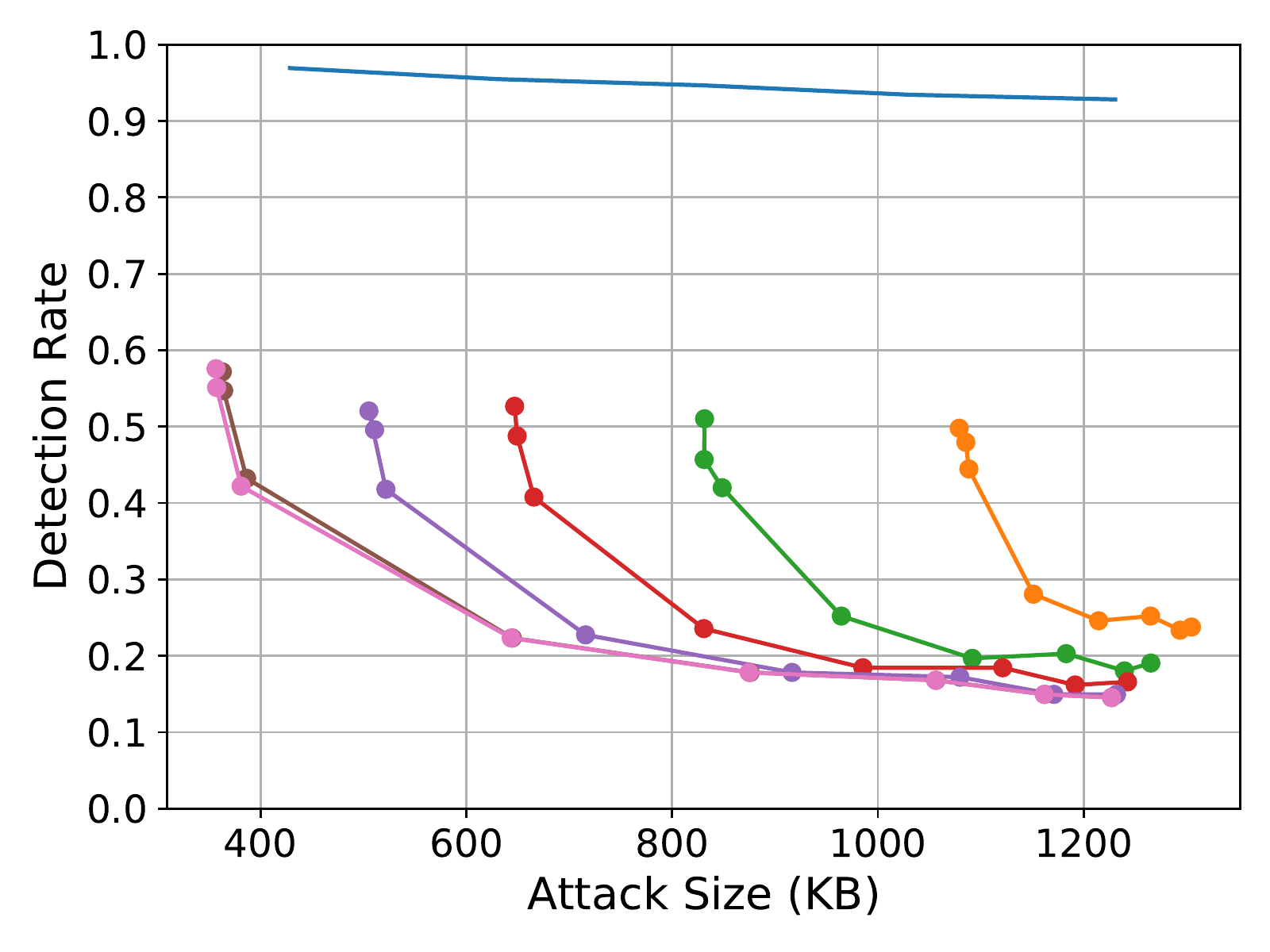}
    \caption{}
    \label{fig:bb_attack}
\end{subfigure}%
\caption{Effectiveness of (a) gradient-based attacks against end-to-end deep networks, and (b) gradient-free attacks against robust decision-tree classifier trained on hand-crafted features. Symbols "E'' and "P'' specifies the dataset used at training time, while percentages describe the average amount of injected bytes with respect to the input size of the network.}
\label{fig:attacks}
\end{figure*}
We now showcase the impact of practical attacks against machine learning Windows malware detectors.
We first explain which tools we leverage for building practical attacks, and then we consider three different experimental settings: (i) gradient-based (white-box) attacks against end-to-end network-based detectors that take in input programs as-is without extracting features, (ii) gradient-free (black-box) attacks against a particular tree-based detector trained on hand-crafted features, and (iii) transfer attacks against online anti-malware commercial products.

\subsubsection{Implementation.}
To craft adversarial malware, we leverage \emph{SecML Malware}~\cite{demetrio2021secmlmalware}, a Python library that implements most of the previously-described optimizers and practical manipulations of Windows programs.
This library is designed to be compliant with the four properties we require to deliver practical attacks against machine learning models.
SecML Malware also has a command-line interface, named \emph{ToucanStrike}, which enables the creation of adversarial examples by typing commands in a shell terminal.
SecML Malware is the basic building block for hosting and collecting adversarial malware attacks, as it can also be easily extended to include novel attacks, while ToucanStrike provides an immediate interface for creating adversarial attacks with no coding skills required.

\subsubsection{Bypassing network-based malware detectors.}
We start by highlighting how end-to-end networks are vulnerable to adversarial attacks in Figure~\ref{fig:wb_attack}.
These networks, trained on either the open source EMBER dataset~\cite{demetrio21-tifs} (E) or on proprietary data (P), take in input programs as strings of bytes without extracting hand-crafted features.
We consider a gradient-based attack coupled with the \emph{extend} manipulation and bound the maximum number of injected bytes (i.e., the perturbation size $\epsilon$) to 4KB. We run the attack for 50 iterations against networks with different architectures, as detailed in~\cite{demetrio21-tops}. The results show that our adversarial malware samples deteriorate the detection rate of the given models in just 5 to 20 iterations of our algorithm.

\subsubsection{Bypassing tree-based malware detectors.} In this case, we apply a gradient-free attack against a popular decision-tree model trained on hand-crafted features, as detailed in~\cite{demetrio21-tifs}.
This attack uses a genetic optimizer to select portions of sections extracted from benign programs and injects such content into new sections created inside the input malware (i.e., performing a \emph{section-injection} attack).
The idea behind injecting content from benign programs is to drastically reduce the number of queries that are normally needed by state-of-the-art black-box attacks to optimize adversarial examples~\cite{demetrio21-tifs}.
The attack can also control the perturbation size $\epsilon$, i.e., the number of injected bytes, by means of a specific penalty added to the loss function.
The results are reported in Figure~\ref{fig:bb_attack},
showing that evasion is achieved by performing very few queries (i.e., from 10 to 500, against the tens of thousands typically requested by state-of-the-art black-box attacks), and with small injected payloads (700 KB on average).

\begin{figure}[h!]
    \centering
    \includegraphics[width=\linewidth]{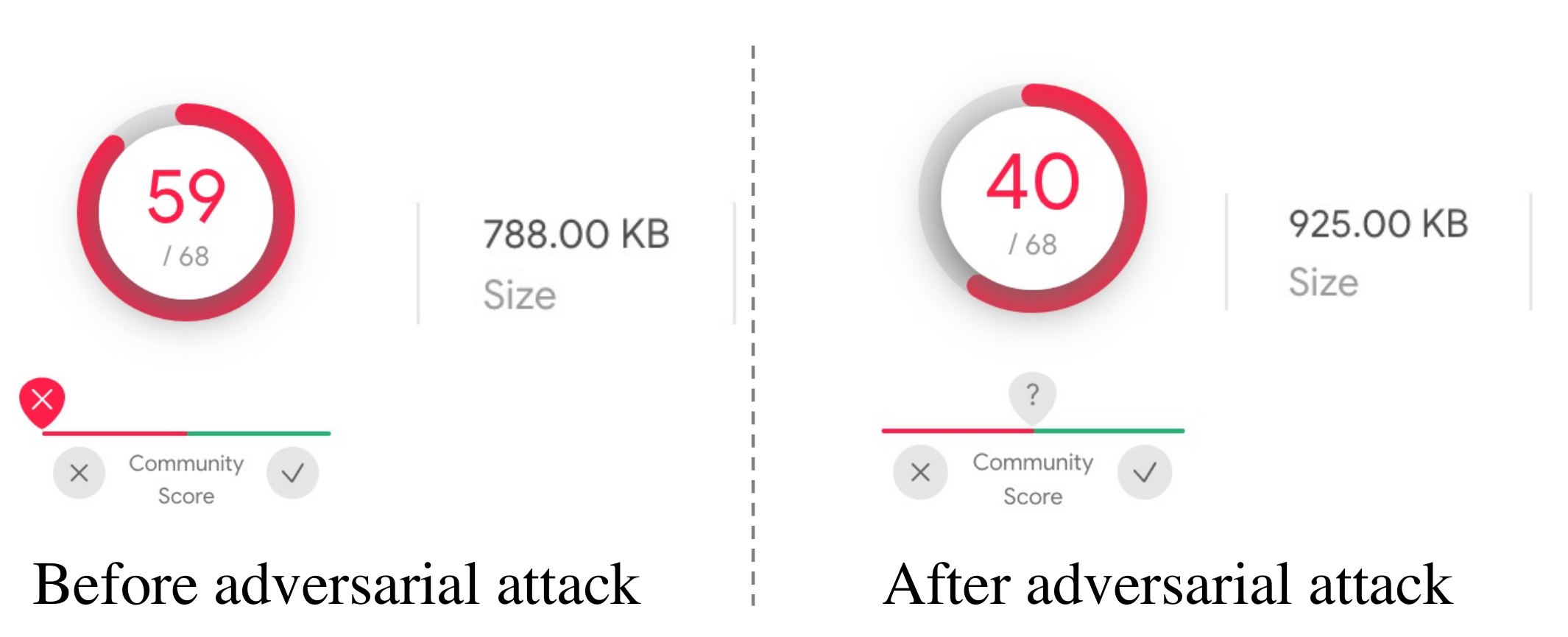}
    \caption{Testing online antivirus hosted on VirusTotal, before and after the application of adversarial noise to a Petya ransomware sample.}
    \label{fig:vt_adv}
\end{figure}
\begin{figure}[h!]
    \centering
    \includegraphics[width=\linewidth]{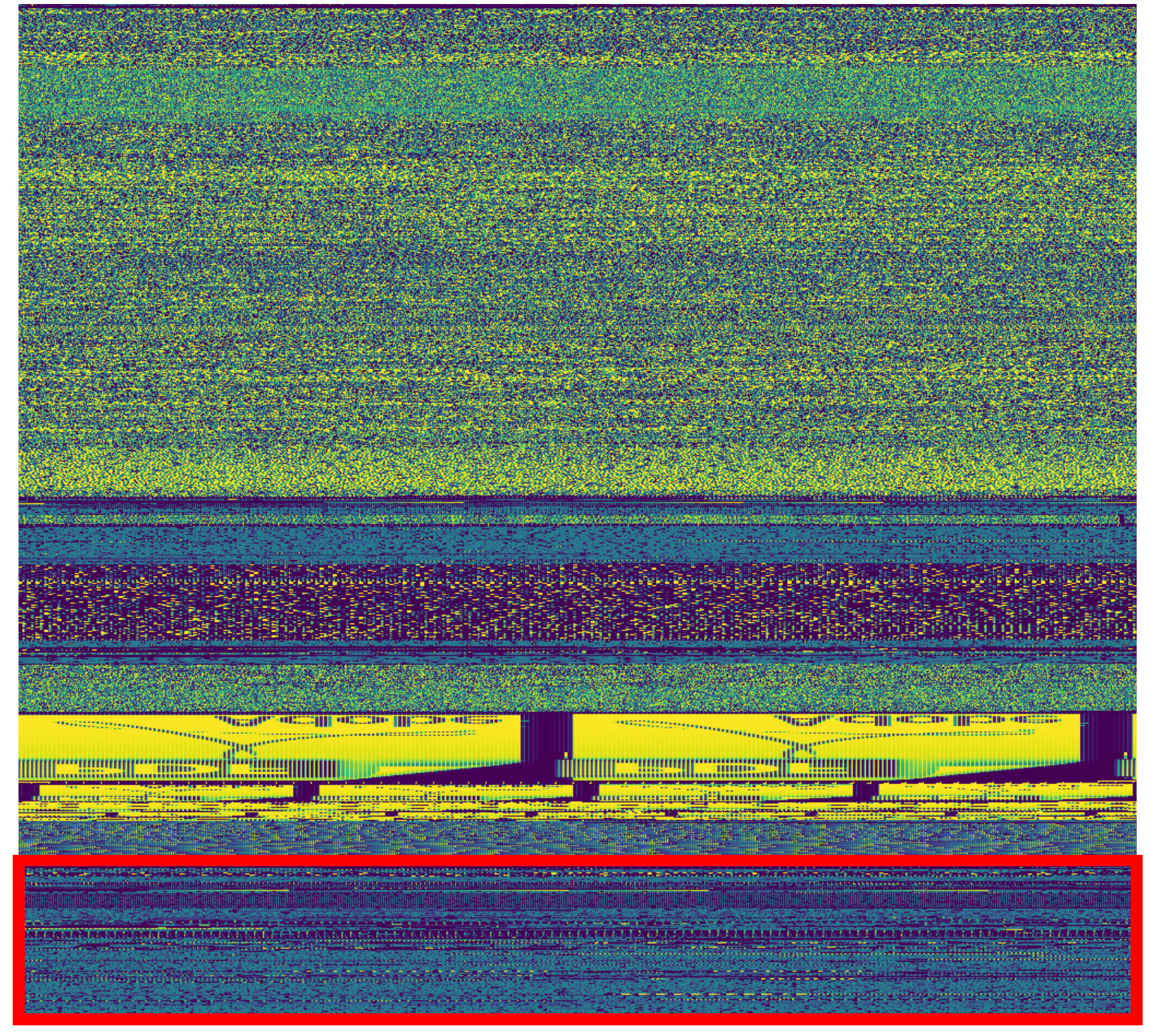}
    \caption{The adversarial version of the Petya ransomware, where we highlight in red the payload added by the section-injection attack.}
    \label{fig:visual_adv}
\end{figure}

\subsubsection{Bypassing commercial anti-malware products.}
We conclude by showing how to create an adversarial version of the infamous Petya ransomware, using the same \emph{section-injection} attack detailed in the previous case~\cite{demetrio21-tifs}. We optimize the adversarial malware example against the decision-tree model described in the previous paragraph and then perform a \emph{transfer attack} by uploading the adversarial malware to \emph{VirusTotal}, which is a popular service that scans the input file with multiple commercial products. We track the number of anti-malware solutions that detect the malware program before and after applying the adversarial manipulations.
As shown in Figure~\ref{fig:vt_adv}, the number of detections decreases from 59 to 40, meaning that 19 commercial products have been evaded with this simple transfer attack.
This example is just one paradigmatic case to convey the intuition of how machine learning malware detectors based on static program analysis can be brittle.
In practice, recent results show that this phenomenon can happen at scale, as the same attack has been demonstrated on many other malware programs, giving the attacker a systematic way for computing slightly-perturbed samples that evade commercial products~\cite{demetrio21-tifs}.
We finally show the adversarial variant of the Petya ransomware computed before in Figure~\ref{fig:visual_adv}, by rendering bytes with different colors and highlighting the injected payload in red.

\section{Conclusions and Future Work}
In this article, we have shown that we can make a step towards a more systematic and scalable attacking methodology for machine learning algorithms, by proposing a framework that mitigates the four issues that hinder the application of attacks in this domain.
This framework consists of two essential building blocks, i.e., the practical manipulations to be defined within the given application-specific constraints, and the optimizer which will be used to fine-tune them.
We have discussed a use case on Windows malware detection, highlighting how one can instantiate our framework to create attacks with ease, and raising an alarm in the field since already-deployed technologies are weak against adversarial attacks.

As future work, we would like to imagine the presence of tools that will help developers and security engineers not only apply adversarial attacks against their models, but also to test, debug, apply version control, perform unit testing, and more.
In an ideal world, we would use an integrated development environment (IDE) similar to the one we use for regular software, where a developer has full access to the same tools they usually use when coding.
This would lead to the formalization of coding patterns and best practices also for machine learning algorithms, and push safety and robustness as a consequence.
Finally, we foresee a thriving environment where also machine learning vulnerabilities are considered as important as the ones discovered in regular programs, since they are already deployed in safety-critical and security-sensitive settings, as the one reported in our empirical analysis.

\section{Acknowledgements}

This work was partly supported by the PRIN 2017 project RexLearn (grant no. 2017TWNMH2), funded by the Italian Ministry of Education, University and Research; and by the TESTABLE project, funded by the European Union's Horizon 2020 research and innovation program (grant no. 101019206).


\begin{thebibliography}{10}

\bibitem{anderson2017evading}
H.~S. Anderson, A.~Kharkar, B.~Filar, and P.~Roth.
\newblock Evading machine learning malware detection.
\newblock {\em Black Hat}, pages 1--6, 2017.


\bibitem{biggio13-ecml}
B.~Biggio, I.~Corona, D.~Maiorca, B.~Nelson, N.~\v{S}rndi\'{c}, P.~Laskov,
  G.~Giacinto, and F.~Roli.
\newblock Evasion attacks against machine learning at test time.
\newblock In H.~Blockeel, K.~Kersting, S.~Nijssen, and F.~\v{Z}elezn\'{y},
  editors, {\em Machine Learning and Knowledge Discovery in Databases (ECML
  PKDD), Part III}, volume 8190 of {\em LNCS}, pages 387--402. Springer Berlin
  Heidelberg, 2013.

\bibitem{biggio18}
B.~Biggio and F.~Roli.
\newblock Wild patterns: Ten years after the rise of adversarial machine
  learning.
\newblock {\em Pattern Recognition}, 84:317--331, 2018.

\bibitem{carlini2018audio}
N.~Carlini and D.~Wagner.
\newblock Audio adversarial examples: Targeted attacks on speech-to-text.
\newblock In {\em 2018 IEEE Security and Privacy Workshops (SPW)}, pages 1--7.
  IEEE, 2018.

\bibitem{dalvi04}
N.~Dalvi, P.~Domingos, Mausam, S.~Sanghai, and D.~Verma.
\newblock Adversarial classification.
\newblock In {\em Tenth ACM SIGKDD Int'l Conf.on Knowledge
  Discovery and Data Mining (KDD)}, pages 99--108, Seattle, 2004.

\bibitem{demetrio2021secmlmalware}
L.~Demetrio and B.~Biggio.
\newblock secml-malware: A python library for adversarial robustness evaluation
  of windows malware classifiers, 2021.

\bibitem{demetrio21-tifs}
L.~Demetrio, B.~Biggio, G.~Lagorio, F.~Roli, and A.~Armando.
\newblock Functionality-preserving black-box optimization of adversarial
  windows malware.
\newblock {\em IEEE Trans. on Information Forensics and Security},
  16:3469--3478, 2021.

\bibitem{demetrio21-tops}
L.~Demetrio, S.~E. Coull, B.~Biggio, G.~Lagorio, A.~Armando, and F.~Roli.
\newblock Adversarial {EXE}mples: A survey and experimental evaluation of
  practical attacks on machine learning for windows malware detection.
\newblock {\em ACM Trans. Priv. Secur.}, 24(4), September 2021.

\bibitem{demontis2017yes}
A.~Demontis, M.~Melis, B.~Biggio, D.~Maiorca, D.~Arp, K.~Rieck, I.~Corona,
  G.~Giacinto, and F.~Roli.
\newblock Yes, machine learning can be more secure! a case study on android
  malware detection.
\newblock {\em IEEE Trans. on Dependable and Secure Computing},
  16(4):711--724, 2017.

\bibitem{grosse2017adversarial}
K.~Grosse, N.~Papernot, P.~Manoharan, M.~Backes, and P.~McDaniel.
\newblock Adversarial examples for malware detection.
\newblock In {\em European Symposium on Research in Computer Security}, pages
  62--79. Springer, 2017.

\bibitem{kreuk2018deceiving}
F.~Kreuk, A.~Barak, S.~Aviv-Reuven, M.~Baruch, B.~Pinkas, and J.~Keshet.
\newblock Deceiving end-to-end deep learning malware detectors using
  adversarial examples.
\newblock {\em Workshop on Security in Mach. Learn. (NeurIPS)}, 2018.

\bibitem{lucas2021malware}
K.~Lucas, M.~Sharif, L.~Bauer, M.~K. Reiter, and S.~Shintre.
\newblock Malware makeover: Breaking ml-based static analysis by modifying
  executable bytes.
\newblock In {\em Proc. of the 2021 ACM Asia Conference on Computer and
  Communications Security}, pages 744--758, 2021.

\bibitem{maiorca19-csur}
D.~Maiorca, B.~Biggio, and G.~Giacinto.
\newblock Towards adversarial malware detection: Lessons learned from
  {PDF}-based attacks.
\newblock {\em {ACM} Comput. Surv.}, 52(4):78:1--78:36, 2019.

\bibitem{nelson08}
B.~Nelson, M.~Barreno, F.~J. Chi, A.~D. Joseph, B.~I.~P. Rubinstein, U.~Saini,
  C.~Sutton, J.~D. Tygar, and K.~Xia.
\newblock Exploiting machine learning to subvert your spam filter.
\newblock In {\em LEET'08: Proc. of the 1st USENIX Workshop on
  Large-Scale Exploits and Emergent Threats}, pages 1--9, Berkeley, CA, USA,
  2008. USENIX Association.

\bibitem{szegedy14-iclr}
C.~Szegedy, W.~Zaremba, I.~Sutskever, J.~Bruna, D.~Erhan, I.~Goodfellow, and
  R.~Fergus.
\newblock Intriguing properties of neural networks.
\newblock In {\em International Conference on Learning Representations}, 2014.

\end{thebibliography}

\begin{IEEEbiography}{Luca Demetrio} is a Postdoctoral Researcher in the Department of Electrical and Electronic Engineering at the University of Cagliari, Italy.
Demetrio received a Ph.D. in computer science from the University of Genova in 2021.
His scientific interests cover the overlapping between adversarial machine learning and computer security, with a strong focus on understanding the weaknesses of malware detectors.
Contact him at luca.demetrio93@unica.it.
\end{IEEEbiography}

\begin{IEEEbiography}{Battista Biggio} is an Assistant Professor in the Department of Electrical and Electronic Engineering at the University of Cagliari, Italy, and co-founder of Pluribus One. He has provided pioneering contributions to the field of machine learning security, being the first to demonstrate gradient-based attacks on machine learning models. He is a Senior Member of the IEEE and a member of the International Association for Pattern Recognition. Contact him at battista.biggio@unica.it.
\end{IEEEbiography}

\begin{IEEEbiography}{Fabio Roli} is a Full Professor of Computer Engineering at the University of Genova, Italy, and Founding Director of the Pattern Recognition and Applications laboratory at the University of Cagliari. He is partner of the company Pluribus One that he co-founded. He has been doing research on the design of pattern recognition and machine learning systems for thirty years, providing seminal contributions to the fields of multiple classifier systems and adversarial machine learning. He has been appointed Fellow of the IEEE and Fellow of the International Association for Pattern Recognition. He is a recipient of the Pierre Devijver Award for his contributions to statistical pattern recognition. Contact him at fabio.roli@unige.it.
\end{IEEEbiography}

\end{document}